# Generative Steganography with Kerckhoffs' Principle

Yan Ke, Min-qing Zhang, Jia Liu, Ting-ting Su, Xiao-yuan Yang

*Key Laboratory of Network and Information Security Under the Chinese People Armed Police Force* (*PAP*).
*Engineering University of PAP, Xi'an, 710086, China.*
15114873390@163.com

***Abstract*—The distortion in steganography that usually comes from the modification or recoding of the cover image during the embedding process. And it is the embedding distortion that leaves the steganalyzer with possible discrimination. Therefore, we propose generative steganography with Kerckhoffs' principle (GSK) in this paper. In GSK, the secret messages are generated by a cover image using a generator rather than embedded into the cover, which results in no modifications to the cover. To ensure security, the generators are trained to meet Kerckhoffs' principle based on generative adversarial networks (GANs). Everything about the GSK system is public knowledge for the receivers, except the extraction key. The secret messages can be outputted by the generator if and only if the extraction key and the cover image are both inputted. In the generator training procedures, there are two GANs (Message-GAN and Cover-GAN) that are designed to work jointly, making the generated results under the control of the extraction key and the cover image. We provide experimental results for the training process. We present an example of the working process by adopting a generator trained on the dataset MNIST, which demonstrates that GSK can use a cover image without any modification to generate messages. Furthermore, only meaningless results would be obtained without the extraction key or the cover image.**

## I. INTRODUCTION

STEGANOGRAPHY sends out secret messages by embedding them into an innocent cover. The goal of steganography is to conceal the hidden channel using the public channel. However, there appear to be two drawbacks that deserve further research, including 1) the insufficient resistibility to steganalysis and 2) the system security that relies on the secrecy of the design or implementation of steganography.

The first drawback is derived from the modification or recoding of covers. There is always a modification or recoding of the cover after embedding, which is designed to appear as innocent as possible to ensure that the hidden channel remains undetected. Such changes in the covers leave the steganalyzer with possible discrimination, since it presents distortion that might be measured by establishing a distortion function. Since the Syndrome-Trellis Codes (STC) was first proposed [1], steganography that minimizes a heuristically-defined embedding distortion has been widely used [2], making the design of distortion function (DF) the only focus. With the evaluation dimensions of DF increasing, the construction getting more complex, and deep-learning algorithms being introduced into steganalysis [3][4], it appears that any change in the cover might present a hidden risk to expose the hidden data.

The other drawback lies in the security of the steganography system. The security (the undetectability of secret messages and the hidden channel) always relies on the secrecy of the design or implementation (*i.e.*, "*security through obscurity*"). However, methods on "*security through obscurity*" may have theoretical or actual security vulnerabilities [5]. In general, the design or implementation of a system is hard to be kept secrecy, since there are too many details that might expose it, and by taking advantage of the insecure aspects of a steganography algorithm, specific steganalysis methods often show more advantages in feasibility than universal methods [6]. In contrast to "*security through obscurity*", Kerckhoffs' principle [7] (a fundamental principle of cryptosystem) is more appropriate for modern security systems. It says that a cryptosystem should be secure even if everything about the system, except the key, is public knowledge. The Fewer and simpler secrets that one must keep make it easier to ensure system security.

Our research attempts to treat the above drawbacks through a distinctive method. We propose Generative Steganography with Kerckhoffs' principle (GSK for short) in this paper. In GSK, the secret messages are obtained by using a cover to generate them rather than recoding the cover to carry them. There is no modification of the cover, thus resulting in no distortion for steganalysis. The generator can be trained based on the emerging technology, generative adversarial networks (GAN) [8]. To ensure security, Kerckhoffs' principle is also introduced. Furon *et al.*, has translated Kerckhoffs' principle from cryptography to data hiding and classified the setups of watermarking attacks into four categories in [9] and [10]. Security levels have been defined in these setups.

Y. Ke (corresponding author), M. Zhang, J. Liu, T. Su, and X. Yang are with the Key Laboratory of Network and Information Security Under the Chinese People Armed Police Force(PAP), Engineering University of PAP, Xi'an, 710086, China (e-mail: 15114873390@163.com; api_zmq@126.com; twinly77@gmail.com; suting 0518@163.com; yxyangyxyang@163.com).

The highest level that is called stego-security is defined in the *Watermarked Only Attack* setup based on Kerckhoffs' principle [11]. Currently, Kerckhoffs' principle has drawn more and more attention [12][13][14]. Our scheme is also designed to meet Kerckhoffs' principle, *i.e.*, the secret messages should be secure even if everything, including the trained generators of GSK, except for the extraction key, are public knowledge. Specifically, the generator is required to output the desired results if and only if the extraction key and the cover image are both inputted.

The remainder of this paper is organized as follows: We detail the GSK framework in the following section. Section III defines the training architecture of the generator. Section IV describes the working processes of GSK. Experiment results are demonstrated in Section V. Section VI concludes this research and details our future work.

## II. Framework of GSK

The exchange of the secret messages in GSK is achieved by the cover images' generating rather than their modification. Therefore, the function of the cover image is to conceal the existence of message exchange and to act as the seed for the GSK generators. To meet Kerckhoffs' principle, the generators are trained with the following two objectives:

1) The secret message can be outputted if and only if both the extraction key and the cover image are inputted.
2) Neither the extraction key itself nor the cover image can reveal any knowledge regarding the secret message.

The GSK framework is shown in Fig. 1. The trained generators would be public knowledge. First, the sender chooses a natural image $\boldsymbol{I}$ as the cover. Then, he can obtain the extraction key $\boldsymbol{k}$ using the secret message $\boldsymbol{s}$ and the cover image $\boldsymbol{I}$ (as shown in Fig. 1a). Next, $\boldsymbol{I}$ is sent out through the public channel and $\boldsymbol{k}$ through the key channel. Similar to [9] and [10], we consider that there is a channel to share the extraction key. We pay no further attention to the key channel (as shown in Fig. 1b).

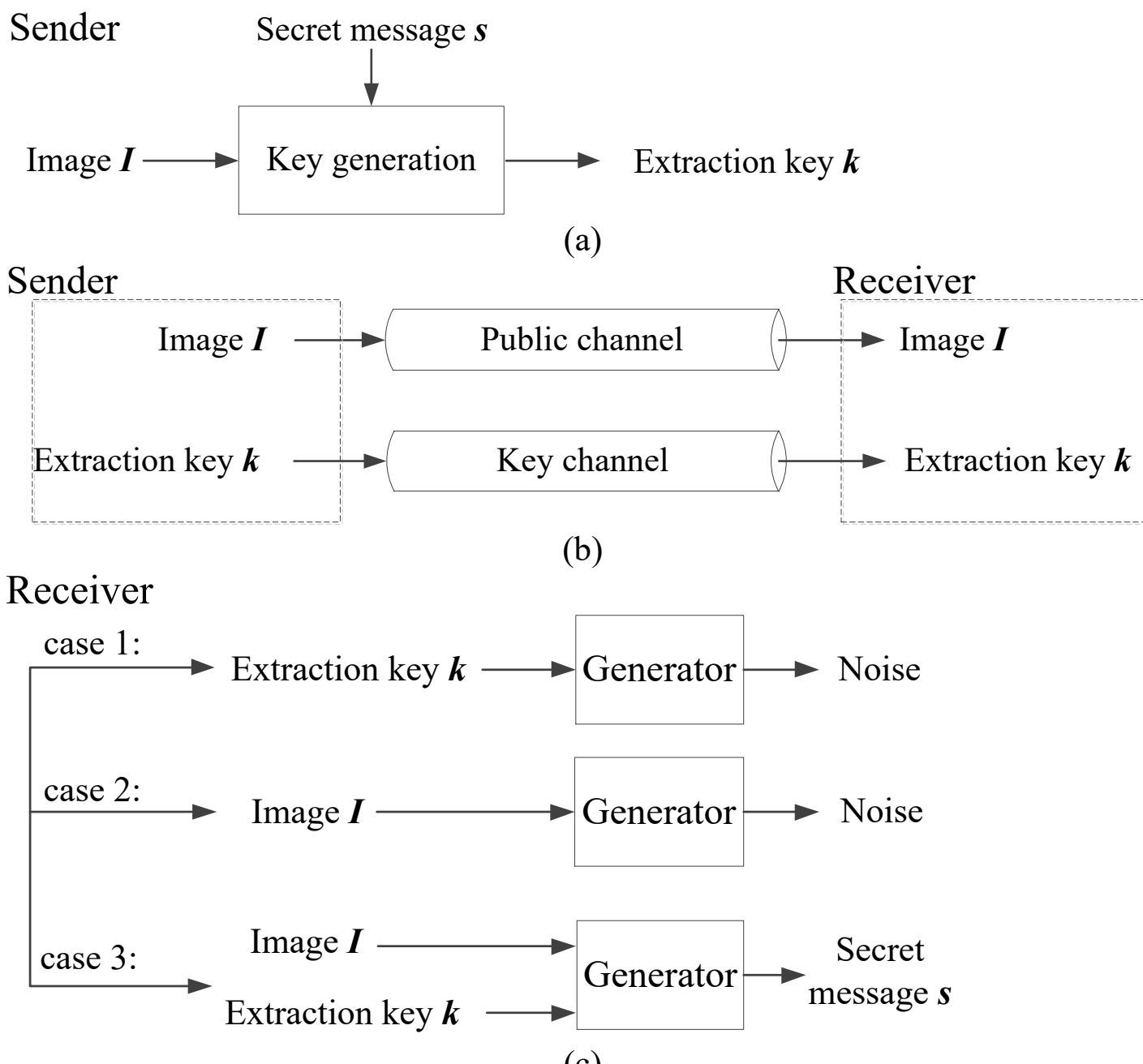

Fig. 1. GSK Framework: (a) Extraction key generation; (b) Data delivery; (c) The three cases of the receivers.

As for the receivers (as shown in Fig.1c), there are 3 cases: In *Case 1*, only $\boldsymbol{k}$ is received, which corresponds to a failed message delivery, and the output is only noise. In *Case 2*, only $\boldsymbol{I}$ is received, which corresponds to an interception from attackers, and the output is only noise (we assume a worst-case scenario here that attackers would intercept the key channel due to its random-like form and $\boldsymbol{I}$ should be undetectable during its lossless delivery). In *Case 3*, $\boldsymbol{I}$ and $\boldsymbol{k}$ are both received, and message $\boldsymbol{s}$ can be obtained.

## III. Training Architecture of the Generator

The generator is the crucial piece of GSK. In this paper, we obtain the generators based on GAN. GAN generates artificial samples that are indiscernible from the real counterparts via the competition between a generator ($G$) and a discriminator ($D$), *i.e.*, alternating the maximization and minimization steps in (1).

$$\min_G \max_D V(D,G) = E_{s\sim}\left[\log D(\boldsymbol{s})\right] + E_{\sim}\left[\log\left(1 - D\left(G(\boldsymbol{x})\right)\right)\right] \tag{1}$$

where $D(\boldsymbol{s})$ is the probability that $\boldsymbol{s}$ is a real image rather than synthetic, and $G(\boldsymbol{x})$ is a synthetic image for input $\boldsymbol{x}$. GAN will finally reach a state of Nash equilibrium of $G$ and $D$. The performances of $G$ and $D$ both get promoted and there is a 50% chance for $D$ to distinguish real samples from generated ones by $G$.

GAN initially works under the unsupervised learning. It needs to input only noise to output a desired result. If so, we cannot control the output with a key. Therefore, our training goal is more complex than the original GAN. We attempt to establish two associations between the message $\boldsymbol{s}$ and the key $\boldsymbol{k}$, and between the message $\boldsymbol{s}$ and the cover $\boldsymbol{I}$. There are two GANs in the training (Message-GAN and Cover-GAN) that are designed to work jointly to place the output under the control of $\boldsymbol{k}$ and $\boldsymbol{I}$.

### *A. Message-GAN*

The objective of Message-GAN is to use feature codes to control the output. Feature codes are a set of discrete random variables that represent the attributes of the samples from some dataset. They are independent and salient in generating meaningful samples, *e.g.*, “face contours”, “skin colors”, and “fat or thin”, *etc.*, are features of the faces from CelebA. “Number values”, “font weight”, *etc.*, are features of the digits from MNIST. As shown in Fig.2, we use noise $\boldsymbol{x}$ with feature codes $\boldsymbol{f}$ as inputs. The generator $G_M$ and the discriminator $D_M$ are adversarial neural networks. $G_M$ aims to generate artificial samples indiscernible from the real samples and to ensure that artificial samples have the features predefined by $\boldsymbol{f}$. Feature codes are equal to the predefined constraint conditions on the output of $G_M$. Ideally, there should be only meaningless outputs without $\boldsymbol{f}$.

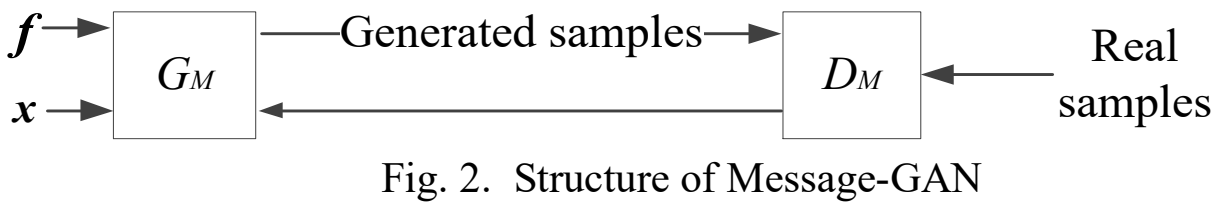

Fig. 2. Structure of Message-GAN

There are several existing GANs that could act as Message- GAN, including InfoGAN [15], Auxiliary Classifier GAN [16], and Conditional GAN [17]. Their performances might vary on different training datasets, which deserves further research. In this paper, we use InfoGAN as Message-GAN, because it can intuitively explain the function of Message-GAN. In InfoGAN, there is high mutual information between $\boldsymbol{f}$ and $G_M(\boldsymbol{x}, \boldsymbol{f})$. The mutual information $I(X; Y)$ between $X$ and $Y$ is the reduction of uncertainty in $X$ when $Y$ is observed:

$$I(X;Y)=H(X)-H(X/Y)=H(Y)-H(Y/X) \tag{2}$$

If $X$ and $Y$ are determinately related, their mutual information gets a maximum. Then, we have the following minimax game:

$$\begin{aligned}\min_{G_M}\max_{D_M} V_M(D_M,G_M) &= E_{s\sim}\left[\log D_M(\boldsymbol{s})\right]+\\ &E_{x\sim}\left[\log\left(1-D_M\left(G_M(\boldsymbol{x},\boldsymbol{f})\right)\right)\right]-\alpha I\left(\boldsymbol{f};G_M(\boldsymbol{x},\boldsymbol{f})\right)\end{aligned} \tag{3}$$

where $\alpha$ is an extra hyperparameter [15]. After training, there will be a high correlation between the outputs of $G_M$ and $\boldsymbol{f}$.

### *B. Cover-GAN*

The objective of Cover-GAN is to make the cover image be the necessary input to the generation of the message. In our framework, the cover image is analogous to the key in the cryptosystem that is necessary for encryption or decryption. Therefore, we have designed Cover-GAN by using the symmetrical encryption GAN in [14]. The structure of Cover-GAN can be abstracted as a communication model in Fig. 3, where $\boldsymbol{z}$ is the necessary input in the communication. A sender wants to transmit some information $\boldsymbol{P}$ to a receiver, and they share the same input $\boldsymbol{z}$. First, the sender inputs $\boldsymbol{P}$ and $\boldsymbol{z}$ into a *generation network* $G_C$ and obtains a random-like output $\boldsymbol{c}$. Then, $\boldsymbol{c}$ is sent out. With $\boldsymbol{c}$, the receiver could recover $\boldsymbol{P}$ by inputting $\boldsymbol{c}$ and $\boldsymbol{z}$ into a *recovery network* $R_C$. There is also an attacker that is attempting to recover $\boldsymbol{P}$ from $\boldsymbol{c}$ without $\boldsymbol{z}$ using an *analysis network* $A_C$.

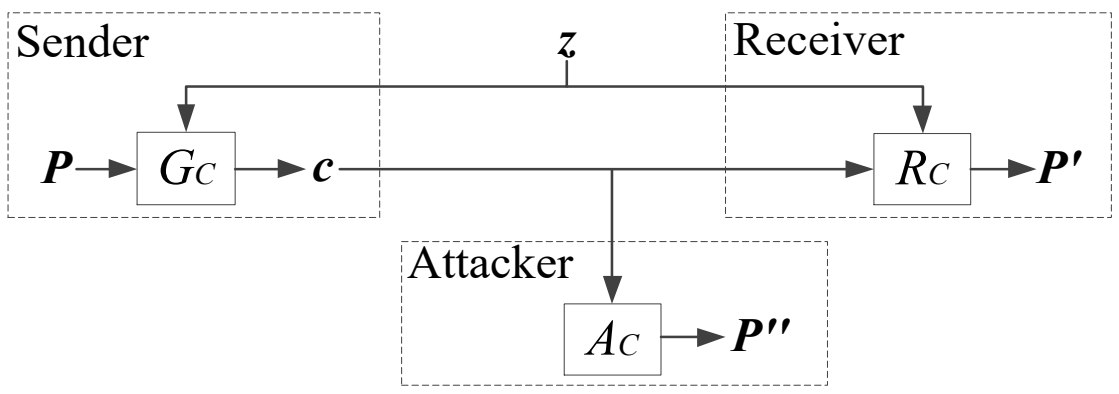

Fig. 3. Structure of Cover-GAN.

$G_C$ and $R_C$ share the same input $\boldsymbol{z}$. The output of $G_C$ is $\boldsymbol{c}$. $R_C$ and $A_C$ can both receive $\boldsymbol{c}$. The two adversarial sides are $A_C$ and ($G_C$, $R_C$). $A_C$ aims to recover $\boldsymbol{P}$. $G_C$ and $R_C$ are jointly trained to ensure that $R_C$ could accurately recover $\boldsymbol{P}$ and that $A_C$ would obtain nothing useful regarding $\boldsymbol{P}$. The output of $R_C$ is $\boldsymbol{P}'$, and the output of $A_C$ is $\boldsymbol{P}''$. According to their different goals, we give the distortion functions of $A_C$ and ($G_C$, $R_C$).

The parameter settings of network $M$ are denoted as $\theta_M$. The L1 distance $d$ on the input $(\boldsymbol{X}, \boldsymbol{X}')$ is introduced to show the distortion between $\boldsymbol{X}$ and $\boldsymbol{X}'$:

$$d(\boldsymbol{X},\boldsymbol{X}')=\sum_{i=1}^{N}\left|x_i-x_i'\right| \tag{4}$$

where $N$ is the length of $\boldsymbol{X}$ and $\boldsymbol{X}'$. The loss functions of $R_C$ and $A_C$ are:

$$L_R(\theta_R) = L_R(\boldsymbol{P}, R(\theta_R, \boldsymbol{z}, \boldsymbol{c})) = E[d(\boldsymbol{P}, \boldsymbol{P}')] \tag{5}$$

$$L_A(\theta_A) = L_A(\boldsymbol{P}, A(\theta_A, \boldsymbol{c})) = E[d(\boldsymbol{P}, \boldsymbol{P}'')] \tag{6}$$

The goal of $A_C$ is to accurately reconstruct $\boldsymbol{P}$, *i.e.*, to minimize distortion between $\boldsymbol{P}$ and $\boldsymbol{P}''$. Thus, the optimal $A_C$ ($O_A$) is obtained:

$$O_A = O_A(\theta_A, \boldsymbol{c}) = \arg\min_{\theta_A}(L_A(\theta_A)) \tag{7}$$

As the adversarial side against $A_C$, networks $G_C$ and $R_C$ aim to successfully communicate and defeat the best possible version of $A_C$. The loss function of ($G_C$, $R_C$) is obtained:

$$L_{GR}(\theta_G, \theta_R) = L_{GR}(L_R(\theta_R), O_A(\theta_A, \boldsymbol{c})) = E[L_R(\theta_R) - O_A(\theta_A, \boldsymbol{c})] \tag{8}$$

where $O_A$ is updated after each training of $A_C$. The optimal $G_C$ and $R_C$ ($O_G$, $O_R$) are obtained by minimizing $L_{GR}(\theta_G, \theta_R)$.

$$(O_G, O_R) = \arg\min_{(\theta_G, \theta_R)}(L_{GR}(\theta_R, \theta_G)) \tag{9}$$

During the training, we start with ($G_C$, $R_C$) to obtain an $R_C$ that could understand $\boldsymbol{c}$ from $G_C$. Then, we fix $G_C$ and $R_C$ and train $A_C$ to obtain an $O_A$. Next, we train $G_C$ and $R_C$ to obtain ($O_G$, $O_R$) with an updated $O_A$. We alternate the training on ($G_C$, $R_C$) and $A_C$ to reach the equilibrium state in which $R_C$ can accurately recover $\boldsymbol{P}$ while $A_C$ has a 50% chance to output the correct bits of $\boldsymbol{P}$. It is noted that the reconstruction error of $A_C$ is not strictly maximized [18]. If it were, all bits of $\boldsymbol{P}''$ would be completely wrong, and $A$ could accurately output $\boldsymbol{P}$ in the next step by flipping all the present bits. Therefore, it is the mutual information between $\boldsymbol{P}''$ and $\boldsymbol{P}$ that gets minimized, and $A_C$ produces answers indistinguishable from random guesses.

## IV. Generative Steganography with Kerckhoffs' Principle

The GSK generators that are trained on different datasets are all public. The scenario of GSK is that a sender intends to send out the secret message $\boldsymbol{s}$ to a receiver by using a cover image $\boldsymbol{I}$ and generators of GSK. There are different generators trained on different datasets. In the application process, the appropriate generator is chosen according to the content type of $\boldsymbol{s}$. Fig. 4 shows the workflow.

The sender's manipulation (Fig. 4a) obtains the extraction key $\boldsymbol{k}$. The sender first chooses the feature codes according to the message $\boldsymbol{s}$ and confirms the feature codes $\boldsymbol{f}$ according to $\boldsymbol{s}$ to ensure that $G_M$ can output $\boldsymbol{s}$ with $\boldsymbol{f}$. Feature choosing module is easy to implement, because the features we can control by Message-GAN are limited, which is discussed in Section V. Then, he chooses random codes $\boldsymbol{f}'$ of the same length as $\boldsymbol{f}$ and calculates $\boldsymbol{f}''$:

$$\boldsymbol{f}'' = \boldsymbol{f} \oplus \boldsymbol{f}' \tag{10}$$

The input of $G_C$ is ($\boldsymbol{P}$, $\boldsymbol{z}$), as described in Section III.B. Here, $\boldsymbol{P}$ is $\boldsymbol{f}''$, and $\boldsymbol{z}$ is the identification code filtered from the cover image $\boldsymbol{I}$ through an advanced appointed method. For simplicity, $\boldsymbol{z}$ is the LSB of $\boldsymbol{I}$ in this paper. The output of $G_C$ is $\boldsymbol{c}$.

The extraction key $\boldsymbol{k}$ is given by ($\boldsymbol{c}$, $\boldsymbol{f}'$). For security reasons, random codes $\boldsymbol{f}'$ instead of feature codes $\boldsymbol{f}$ are used to construct the extraction key $\boldsymbol{k}$ in case $\boldsymbol{k}$ is intercepted and $\boldsymbol{k}$ itself should not divulge any useful information about $\boldsymbol{f}$. Then the cover image $\boldsymbol{I}$ and the extraction key $\boldsymbol{k}$ are sent out as shown in Fig. 1(b).

The receiver's manipulation (Fig. 4b) obtains the secret message $\boldsymbol{s}$ with $\boldsymbol{k}$($\boldsymbol{c}$, $\boldsymbol{f}'$) and $\boldsymbol{z}$ from $\boldsymbol{I}$. First, $\boldsymbol{z}$ and $\boldsymbol{c}$ are inputted into $R_C$ to obtain $\boldsymbol{f}''$. Then $\boldsymbol{f}$ is calculated:

$$\boldsymbol{f} = \boldsymbol{f}'' \oplus \boldsymbol{f}' \tag{11}$$

The secret message $\boldsymbol{s}$ can be outputted by $G_M$ with input $\boldsymbol{f}$. To summarize, $\boldsymbol{k}$ and $\boldsymbol{z}$ are both necessary for the generation of $\boldsymbol{s}$. The security of GSK depends solely on the secrecy of $\boldsymbol{k}$ and the randomizer of $\boldsymbol{f}'$.

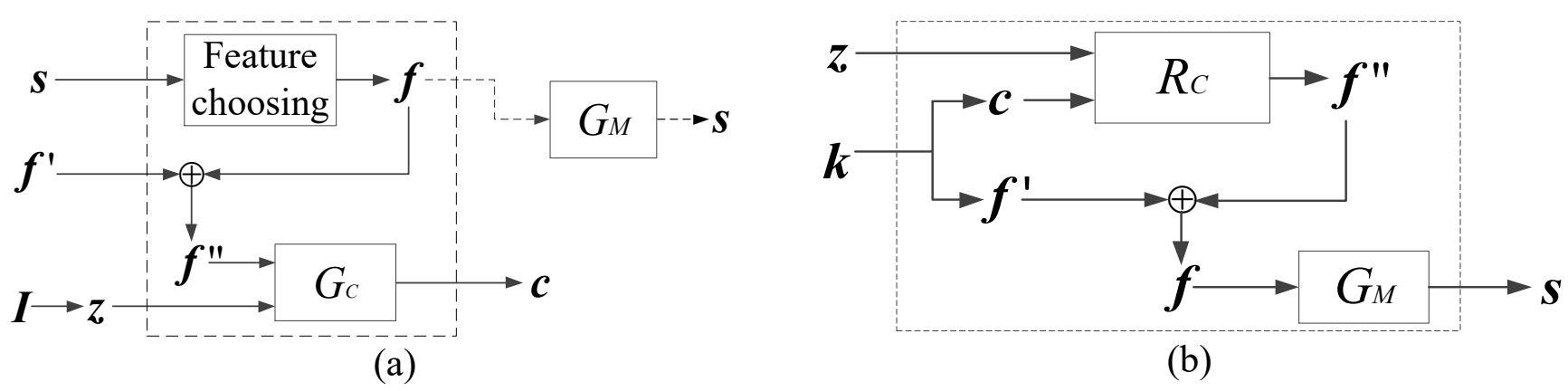

Fig. 4. Workflow: (a) Sender's manipulation; (b) Receiver's manipulation.

## V. Experimental Results and Analysis

First, we demonstrate the experimental results of the training processes including Message-GAN and Cover-GAN. Then, an example is provided to illustrate the working processes of GSK. We plan to release the source code for the experiment on GitHub.

### A. Training processes of GSK

The training of Message-GAN obtains the feature codes ***f*** of a target message. The experiments are based on two datasets, MNIST and CelebA. Secret messages are contained in the output images of the generators. The content types of the messages include decimal handwritten digits (on MNIST) or faces (on CelebA). The training of Cover-GAN ensures that $R_C$ can recover ***f'*** without loss while $A_C$ only outputs random bits.

***Message-GAN***: First, we test on MNIST by mechanically applying the setup in [15]. We use 1 ten-dimensional feature code, 2 latent codes and 62 noise variables as the inputs with a dimension of 74, i.e., the input of the generator $G_M \in \mathbb{R}$ ; The inputs first go into a full connection (FC) layer, followed by the Batch Normalization layer and 1024 Rectified Linear Units (ReLU); Then they go to another FC layer and batch normalization layer with 7×7×128 ReLU; Next, there is a 4×4 convolutional layer with 64 ReLU. The discriminator $D_M$ consists of two nets, the first one of which is to ensure the artificial samples indiscernible from the real samples, and the other is to ensure that artificial samples have the predefined features. The two nets have the same architecture of a 28 ×28 input followed by two 4×4 convolutional layers and another FC layer with 1024 Leaky-ReLU. The output of the first net is a FC layer while the output of the second is a FC layer with a 128-bit batch normalization and LReLU layer.

Feature code ***f*** with length of 4-bit indicates the number of one decimal digit. The outputs are a 28×28 8-bit grey image. The generated digit is recognizable from the output image, such as digit “5” in Fig. 5a. The results in Figs. 5b-5e demonstrate that we can control the number of the generated digit using ***f***, and the digits are randomly generated without ***f*** (Fig.5e). According to [15], more characteristics, such as the rotation direction or the font weight of the written digit, have already been controlled by the feature codes. That could enhance the amount of the meaningful information contained in an output image.

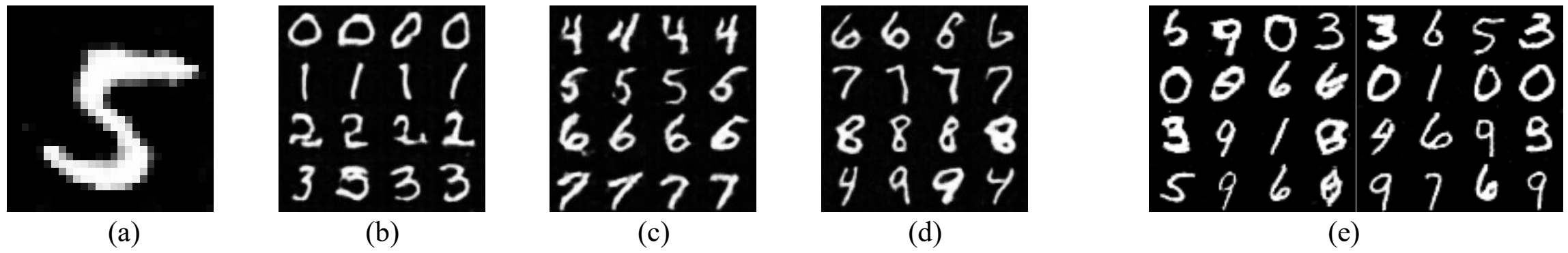

(a) (b) (c) (d) (e)

Fig. 5. Sample results on MNIST: (a) Digit “5” in the output image; (b) ***f***: digits 0,1,2,3; (c) ***f***: digits 4,5,6,7; (d) ***f***: digits 6,7,8,9; (e) Outputs without ***f***.

We continue our test on CelebA to output faces using 10 ten-dimensional feature codes and 128 noise variables as inputs with a dimension of 228. ***f*** here is used to control some highly semantic variations that a generated face carries (*e.g.*, with glasses or not in Fig. 6a and/or with smile or not in Fig. 6b). The specific semantic information in the faces is the message that we can generate. Without ***f***, we use deep convolutional-GAN (DC-GAN) [19] to act as the attacker that might have access to our training dataset but know nothing about ***f***. The outputs of DC-GAN with $3\times10^4$ training steps on CelebA are randomly generated faces that cannot carry any artificially added semantic information as shown in Fig. 6c.

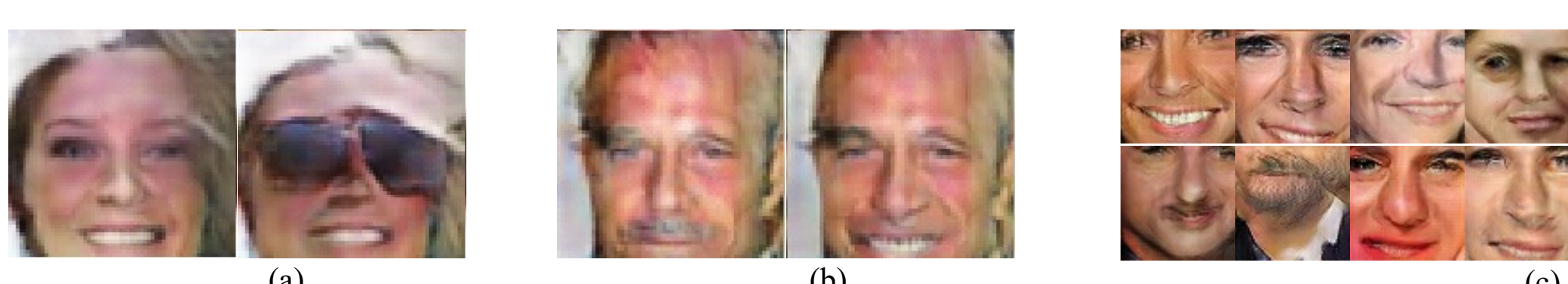

(a) (b) (c)

Fig. 6. Sample results on CelebA: (a) ***f***: with glasses or without glasses; (b) ***f***: with smile or without smile; (c) Outputs of DC-GAN without ***f***.

The experimental results on MNIST or CelebA demonstrate that ***f*** can act as the preconditions for generating the desired messages. However, ***f*** on CelebA cannot be set long yet, and its depiction on human faces is not rich or precise enough at present. Therefore, we use the generator trained on MNIST for experiments in the following sections.

***Cover-GAN***: The lengths of ***P***, ***z***, and ***c*** are all $N$ bits [18] for $N$=16, 32, and 64. $G_C$ concatenates two $N$-bit inputs (***P*** and ***z***) into a $2N$-entry vector. This vector is processed through a $2N\times2N$ fully connected (FC) layer, and then sent through four succession 1-D convolutional layers. The network $R_C$ is identical to $G_C$. $A_C$ takes only ***c*** as the input, and thus has an $N\times2N$ FC layer. We arbitrarily choose a natural image as cover image ***I***. The identification code ***z*** is obtained from the LSB of ***I***.

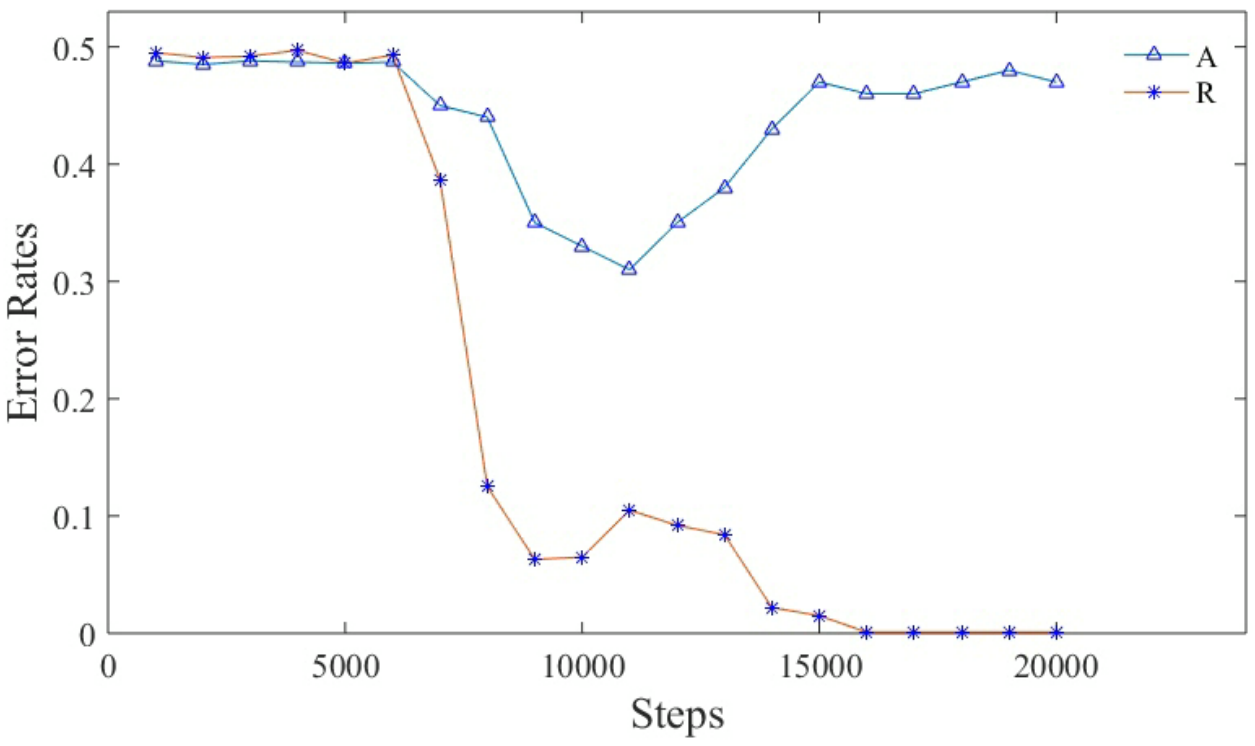

Fig. 7. Error rates of *R* and *A*.

During training, the error rates of $R_C$ and $A_C$ for recovering $\boldsymbol{f''}$ are recoded once per 1000 steps in Fig. 7. After approximately $1.6\times10^4$ steps, $R_C$ can accurately recovery each bit of $\boldsymbol{f''}$ with $\boldsymbol{z}$ while the error rates of $A_C$ approach 50% and then stabilize without $\boldsymbol{z}$. It demonstrates that $\boldsymbol{f''}$ cannot be obtained without $\boldsymbol{I}$, thus ensuring the secrecy of feature codes $\boldsymbol{f}$ and message $\boldsymbol{s}$.

### B. *Working processes of GSK*

We intend to send decimal digits as an example of GSK in this section. To send one decimal digit of 4 bits, we need the trained generators on MNIST and a 4-bit identification code $\boldsymbol{z}$ from the cover image $\boldsymbol{I}$. Then, we obtain the extraction key $\boldsymbol{k}$ of 8 bits. $\boldsymbol{f''}$, $\boldsymbol{z}$ and $\boldsymbol{c}$ all have the same length as $\boldsymbol{f}$. The working processes of GSK are as shown in Fig.8, we choose Lena as the cover image (Fig. 8a) to send digits 0-7 in order. And then the LSBs of Lena are extracted as the code $\boldsymbol{z}$ (Fig. 8c) to obtain the key $\boldsymbol{k}$. The image Lena and $\boldsymbol{k}$ are delivered to the receiver. The digits can be generated in order by using the Lena and $\boldsymbol{k}$ while only random digits can be outputted without the image Lena or $\boldsymbol{k}$ (Fig. 8d-8f).

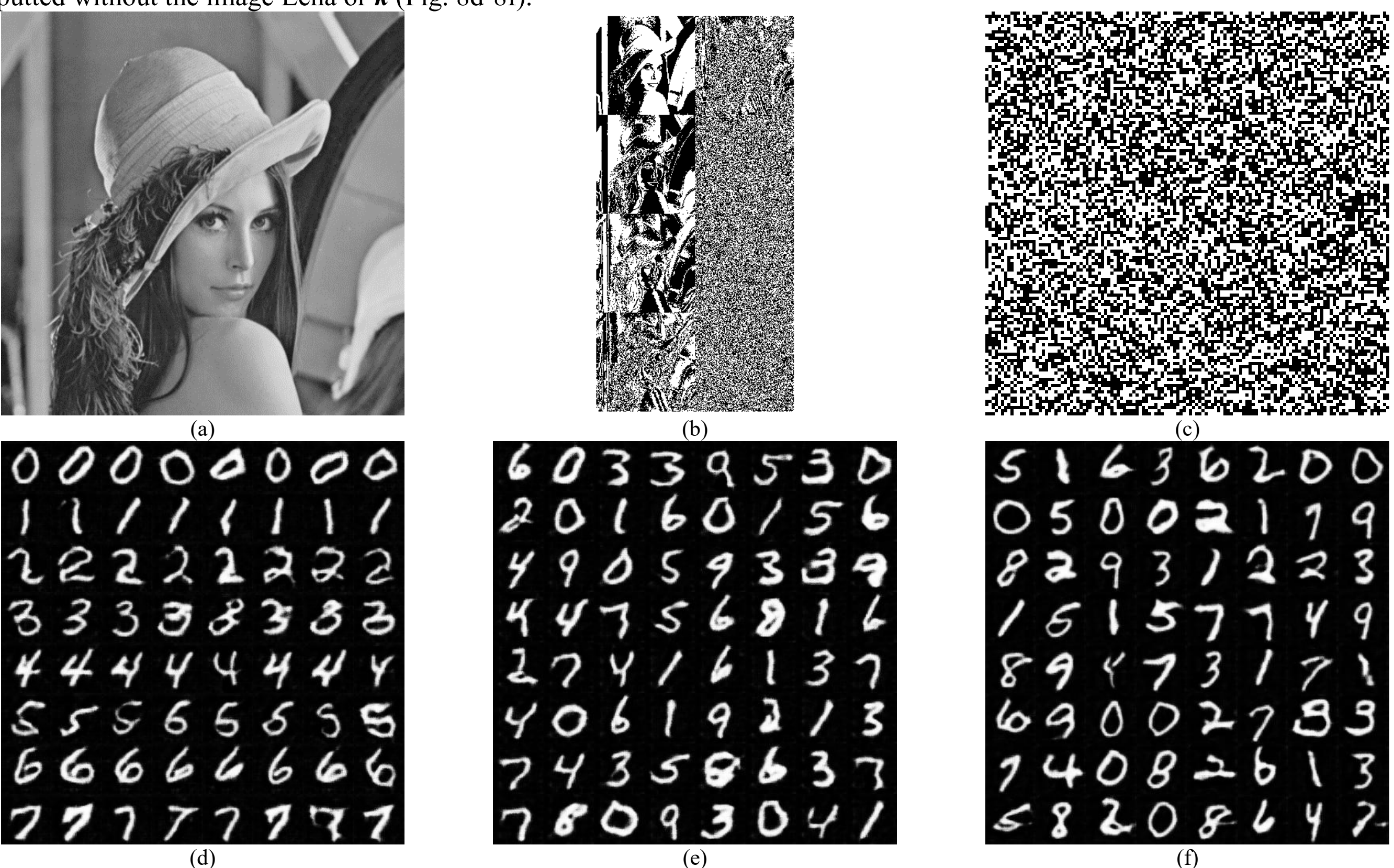

Fig. 8. Working processes of GSK: (a) Cover image Lena; (b) The 8 bit-planes of Lena; (c) LSB plane of Lena; (d) The generated messages with both key and the image Lena; (e) The generated messages without key; (f) The generated messages without the image Lena.

We test the performances of GSK in the 3 cases described in the GSK framework (Fig.1c) using 10 data samples. Each sample has 300 decimal digits as secret messages. The bit-errors between $\boldsymbol{s}$ and the output bits are recorded, and the results of one of our samples are shown in Fig. 9. In *case1* or *case2*, only random digits are generated due to the absence of $\boldsymbol{k}$ or $\boldsymbol{I}$. In *case3*, message $\boldsymbol{s}$ can accurately be obtained.

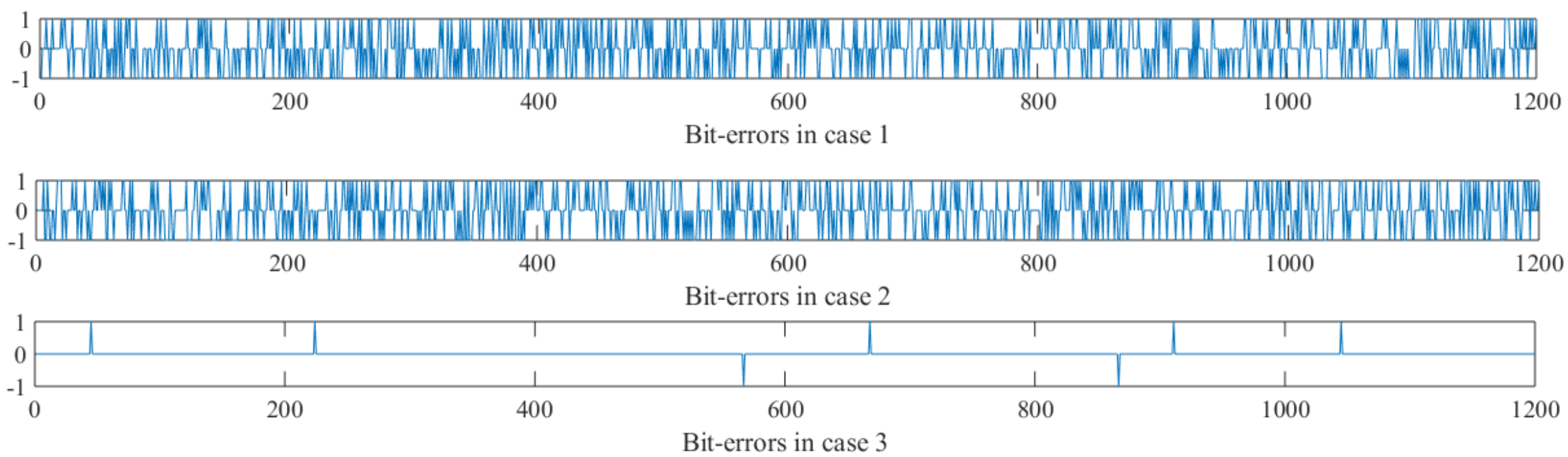

Fig. 9. Bit-errors of the 3 cases.

The bit-error rates between $\boldsymbol{s}$ and the output bits in the 3 cases are calculated. The five best-performing samples are shown in Table. I. The error rates in case 1 or case 2 are close to 0.5 while the error rates in case 3 are close to 0.

TABLE. I
ERROR RATES BETWEEN THE MESSAGE BITS AND THE OUTPUT BITS

| | Sample1 | Sample2 | Sample3 | Sample4 | Sample5 |
|---|---|---|---|---|---|
| Case 1 | 0.5058 | 0.4833 | 0.4908 | 0.4917 | 0.4933 |
| Case 2 | 0.4817 | 0.5125 | 0.4842 | 0.4775 | 0.5050 |
| Case 3 | 0.0042 | 0.0067 | 0.0008 | 0.0017 | 0.0025 |

The ***capacity*** of GSK is determined by the complexity of the training datasets and the control effect of the feature codes. In this paper, one output of the generator on MNIST is a 28×28 8-bit grey image. The theoretical maximum of the capacity is 6272 bits when the output image is fully utilized. In our example, only one digit is carried by an output image. The available information has only 4 bits, and the availability factor (AF) of the output image is 0.0638%.

The ***key size factor*** (KSF) is the length ratio of $\boldsymbol{k}$ and $\boldsymbol{s}$, which indicates the cost of key $\boldsymbol{k}$ for generating message $\boldsymbol{s}$:

$$\mathrm{KSF} = length\ (\boldsymbol{k})\ /\ length(\boldsymbol{s}) \tag{12}$$

The KSF of GSK on MNIST is 2 in our example.

Such a steganography scheme can realize secret communication by exchanging a natural image and the keys, which is somehow similar to cryptography. Their difference lies in the undetectable part (the natural image) of the communication system in GSK, which is the technical goal of steganography, while cryptography transfers secret messages by exchanging two random-like data (ciphertext and key), whose random form are sensitive for the attackers. From this point of view, GSK combines some technical features of cryptography and steganography.

In GSK, the extraction key $\boldsymbol{k}$ consists of $\boldsymbol{f'}$ and $\boldsymbol{c}$. $\boldsymbol{c}$ must be transmitted through the key channel, which might limit the applications of GSK. However, the amount of $\boldsymbol{c}$ is small compared with the total amount of the output image. Further studies should focus on increasing AF and reducing KSF to improve the practicality. Here, the security requires that the key $\boldsymbol{k}$ and the cover image cannot be obtained by the attacker at the same time. Therefore, $\boldsymbol{k}$ is not required to be transmitted through the secret channel but any another channel separated from the cover image, because its content is in the random form and does not reveal any information about the secret message content.

## VI. CONCLUSION

The main contribution of this paper is that we propose a framework of generative steganography with Kerckhoffs' principle. To ensure the security, the generator is required to output the secret message if and only if the cover image and the extraction key are both inputted. To summarize, the exchange of the messages in GSK has been transformed into the exchange of random-like keys and innocent covers. We provide the experimental results to demonstrate the feasibility of GSK and discuss some evaluation criteria of GSK. Future work should mainly focus on 1) enhancing the capacity by applying GSK to the complicated datasets, intensifying the control effect of the feature codes, and 2) optimizing the architecture of Cover-GAN to reduce the KSF of GSK.

## VII. ACKNOWLEDGEMENTS

This work was supported in part by the National Key R&D Program of China under Grant 2017YFB0802000.